\documentclass[journal]{IEEEtran}

\usepackage[colorlinks,citecolor=blue]{hyperref}
\usepackage{graphicx}        
\usepackage{amsmath,amsfonts,amssymb}

\newcommand\eqnref[1]{(\ref{#1})}
\newcommand\figref[1]{Fig.~\ref{#1}}

\newcommand\sectref[1]{Section~\ref{#1}}

\newcommand{\iu}   {j}     

\newcommand{\etal}   {\emph{et al. }}
\newcommand{\bfn}   {\mathbf{n}}
\newcommand{\bfk}   {\mathbf{k}}
\newcommand{\bfm}   {\mathbf{m}}
\newcommand{\bfr}   {\mathbf{r}}
\newcommand{\epscyl}   {\epsilon_{\mathrm{cyl}}}
\newcommand{\kcyl}   {k_{\mathrm{cyl}}}

\newcommand{\figurewidth}  {0.99\linewidth}

\begin{document}
%
\title{A Singularity-free Boundary Equation Method\\
for Wave Scattering}

\author{\authorblockN{Igor Tsukerman}\\
\authorblockA{Department of Electrical and Computer Engineering, The
University of Akron, OH 44325-3904\\
Email: igor@uakron.edu}}


\maketitle

\begin{abstract}
Traditional boundary integral methods suffer from the singularity of Green's kernels.
The paper develops, for a model problem of 2D scattering as an illustrative example,
singularity-free boundary difference equations.
Instead of converting Maxwell's system into an integral boundary form first and discretizing second,
here the differential equations are first discretized on a regular grid and then converted to
boundary difference equations. The procedure involves nonsingular Green's functions on a lattice
rather than their singular continuous counterparts. Numerical examples demonstrate the effectiveness, accuracy
and convergence of the method. It can be generalized to 3D problems and to other classes
of linear problems, including acoustics and elasticity.
\end{abstract}


\IEEEpeerreviewmaketitle

\section{Introduction}
\label{sec:introduction}
%
Boundary equation methods have a long history, with practical applications
dating back to the 1960s. An interesting historical account given by Cheng \& Cheng
\cite{Cheng-Cheng05} includes the work on wave scattering and radiation in 1962--1967 by Friedman \& Shaw,
Chen \& Schweikert, Banaugh \& Goldsmith, Mitzner, and others
\cite{Friedman62}--\cite{Mitzner67}.
In eletromagnetics, boundary integral techniques became very popular due primarily to Harrington's
work published in 1967--68 \cite{Harrington67,Harrington93} (see also
\cite{Tozoni74}--\cite{Bonnet99}).

In traditional boundary integral methods, linear boundary value problems of field analysis
are transformed into integral equations with respect to equivalent sources residing
on the boundaries. In the simplest example of capacitance calculation \cite{Harrington67,Harrington93},
the distributed charge density on conducting plates becomes the principal unknown variable.
By equating the Coulomb potential of that charge to the given potential of the
conductors, one obtains an integral equation. It can then be discretized using variational techniques
(moment methods), collocation and Galerkin methods being particular cases of those.

As all numerical methods, boundary integral techniques do carry some trade-offs.
Their key advantage is the lower dimensionality of the problem: 3D analysis
is reduced to equivalent sources on 2D boundaries and 2D analysis -- to 1D contours.
Another advantage is a natural treatment of unbounded problems (e.g. wave scattering and radiation),
without the artificial domain truncation unavoidable in differential methods such as
finite difference (FD) schemes and the Finite Element Method (FEM).

Integral equation methods have, in general, two major disadvantages. First, the matrices
of the discrete system are almost always full. This is due to the fact that a source
at any point on the boundary contributes to the field at all other points. In contrast,
FD and FE matrices are sparse, with very efficient system solvers available
(iterative: multilevel methods, incomplete factorization and other effective preconditioners;
direct: minimum degree, nested dissection and others; see e.g. \cite{Tsukerman-book07} and references there).
Cases where Green's functions decay rapidly in space, giving rise to quasi-sparse integral equations,
are exceptional (e.g. periodic structures in the electromagnetic band gap regime \cite{Pissoort07}).

Another disadvantage is that the integral kernels in field analysis are singular.
At the surface points, the kernel singularity can usually be handled analytically,
and the fields remain bounded as long as the surfaces are smooth.
However, for points in the vicinity of the surface, the evaluation of the integral is problematic,
as analytical expressions are usually unavailable and numerical quadratures require
extreme care. The same is true for two adjacent surfaces with a narrow gap in between.

Significant progress in Fast Multipole Methods (FMM) \cite{Greengard87,Cheng99,Martinsson-PhD02,Ying04,Fong09}
has helped to alleviate the first disadvantage of boundary methods. FMM accelerates the computation of fields
due to distributed sources -- or equivalently, matrix-vector multiplications for the dense system matrices.

The second disadvantage is more difficult to overcome. Singular kernels are inherent in boundary integral methods
because the fields of point sources are unbounded. However, a drastic change
in the computational procedure leads to a singularity-free method; this is accomplished
by reversing the sequence of stages in the boundary techniques. The standard sequence is

\vskip 0.1in
\noindent
{\tt Differential formulation $\Longrightarrow$ Boundary integral formulation $\Longrightarrow$ Discretization}
\vskip 0.1in

\noindent
The alternative sequence is

\vskip 0.1in
\noindent
{\tt Differential formulation $\Longrightarrow$ Discretization $\Longrightarrow$  Boundary difference problem}
\vskip 0.1in

Discretization of the differential problem is performed on a regular grid and yields an FD scheme.
This scheme is converted -- as explained in the remainder of the paper --
to a boundary problem that involves \emph{discrete} fundamental solutions (Green's functions)
on the grid. Discrete Green's functions, unlike their continuous counterparts, are always
nonsingular.

This general idea is not new. In fact, there are two related but independently developed methodologies
for boundary difference equations. The first one, put forward and thoroughly studied
by Ryaben'kii, Reznik, Tsynkov and others \cite{Ryabenkii02,Reznik82,Tsynkov03,Tsynkov-private}, is known
as the \emph{method of difference potentials} and can be viewed as a discrete analog of the Calderon projection operators in functional analysis \cite{Ryabenkii02}.

The second methodology, called boundary algebraic equations by Martinsson \& Rodin \cite{Martinsson09},
is at least 50 years old (Saltzer \cite{Saltzer58}) and
is a discrete analog of first- or second-order Fredholm boundary integral equations
for potential problems \cite{Martinsson09}.

In comparison with \cite{Martinsson09}, the method of this paper
has several novel features. First, the paper deals -- to my knowledge, for the first time -- with boundary difference equations for electromagnetic wave scattering. In \cite{Martinsson09}, a simple model problem
is considered: the Laplace equation (e.g. electrostatics or heat transfer) in a homogeneous domain
with Dirichlet boundary conditions; the focus of \cite{Martinsson09} is on the mathematical analysis of the respective
boundary difference operators, their spectral properties and the appropriate iterative solvers.

One key distinction between the methodology of Ryaben'kii \cite{Ryabenkii02} and this paper's
is in the choice of the main unknown: the boundary field / potential (Ryaben'kii)
vs discrete sources on the boundary (the present paper).
The treatment via sources parallels that of the continuous boundary integral method
\cite{Harrington93,Tozoni74,Chew07,Peterson97,Bonnet99} and should therefore be intuitive
to applied scientists and practitioners. Further analysis and comparison of these methodologies
will be presented elsewhere. 

An additional novelty of this paper is the use of high-order Flexible Local Approximation
MEthods (FLAME, see \sectref{sec:FLAME}) in the context of boundary difference equations.
Also, this is the first application of FLAME to a 2D boundary
of a generic shape; this is done by approximating this boundary locally
by its osculating circle at any given point.
%
\section{Boundary Difference Equations for a Model Problem}
\label{sec:BDE-model-problem}
%
\subsection{Formulation and Setup}
To fix ideas and explore the potential of the proposed approach, let us consider the classical
2D case of electromagnetic wave scattering as a model problem. It should be emphasized from the outset
that the method has a much broader range of applicability; possible generalizations
are discussed in \sectref{sec:Future-directions}.

\begin{figure}
\centering
  \includegraphics[width=0.6\linewidth]{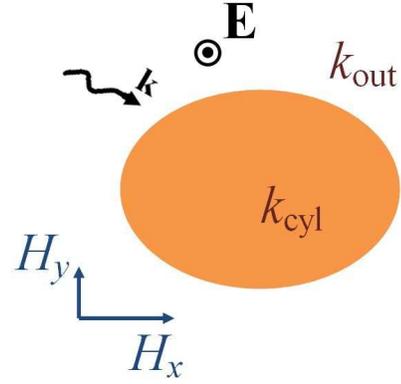}\\
  \caption{Setup of the scattering problem for the $E$-mode.}
  \label{fig:setup-scattering-E-mode}
\end{figure}

Consider a plane wave impinging from the air on a dielectric cylinder (\figref{fig:setup-scattering-E-mode})
with a given dielectric permittivity $\epscyl$. The cross-section of the cylinder could be arbitrary,
but for the sake of simplicity we shall assume that its surface is smooth (no edges or corners).

For definiteness, let us focus on the $E$-mode (TM- or $s$-mode) governed by the familiar equation
for the electric field $\mathbf{E}$ with a single $z$-component:
\begin{equation}\label{eqn:del2-E-k2-E}
    \nabla^2 E(x, y) \,+\, k^2(x, y) E(x, y) ~=~ 0,
    ~~~ k^2 = \omega^2 \mu \epsilon
\end{equation}
where the standard notation for the angular frequency $\omega$, the magnetic permeability $\mu$,
the dielectric permittivity $\epsilon$ and the wavenumber $k$ is used ($k$ is equal to $\kcyl$ inside the scatterer
and to $k_{\mathrm{out}}$ outside). Equation \eqnref{eqn:del2-E-k2-E} should be supplemented by
the standard radiation boundary conditions for the scattered field $E_s = E - E_{\mathrm{inc}}$ at infinity.
The incident field is a plane wave
\begin{equation}\label{eqn:Einc-plane-wave}
    E_{\mathrm{inc}} ~=~ E_0 \exp(-j \bfk_{\mathrm{out}} \cdot \bfr),
    ~~~ \bfr \equiv (x, y)
\end{equation}
where the $\exp(+j \omega t)$ convention for complex phasors is implied.

In a departure from the boundary integral methodology, we now proceed, prior to formulating
a problem on the boundary of the scatterer, to FD discretization. To this end,
let us introduce an \emph{infinite} lattice with a grid size $h$, for simplicity
the same in both $x$ and $y$ directions. Although infinite lattices are not a very common
computational tool, they were already featured prominently in Martinsson \& Rodin's work
\cite{Martinsson09,Martinsson02} as well as in the much earlier report by Saltzer \cite{Saltzer58}.
The actual computation, clearly, never involves an infinite amount of data on the lattice;
in fact, the unknowns are ultimately confined only to the boundaries.

As an auxiliary device, we need to consider the wave equation \eqnref{eqn:del2-E-k2-E}
in the \emph{homogeneous} space with a \emph{constant} generic parameter $k$. Various FD discretizations
of this equation are available; see e.g. Harari's review \cite{Harari06} for further
information and references. Here we settle for the simplest five-point scheme
$$
    \mathcal{L} (h, k) E ~\equiv~
    E(m_x-1, m_y) + E(m_x+1, m_y) - 4 E(m_x, m_y)
$$
\begin{equation}\label{eqn:5-pt-schem-E-mode-const-k}
    + E(m_x, m_y-1) + E(m_x, m_y+1)
    ~+~ k^2 h^2 E(m_x, m_y) ~=~ 0
\end{equation}
where $E(m_x, m_y)$ is the field value at a grid point characterized by an integer double index
$\bfm \equiv (m_x, m_y) \in \mathbb{Z}^2$. As reflected in the notation, the coefficients of the difference
operator $\mathcal{L}$ depend on the mesh size and on the wavenumber; this may not be
explicitly indicated if there is no possibility of confusion.

Associated with $\mathcal{L}$ is its Green's function $g(m_x, m_y)$
defined as the solution of
\begin{equation}\label{eqn:Lg-eq-delta}
    \mathcal{L} g ~=~ \delta
\end{equation}
with the boundary condition
\begin{equation}\label{eqn:g-eq-G}
    g(m_x, m_y; k) \rightarrow G(m_x h, m_y h; k) ~~\mathrm{as} ~ (m_x, m_y) \rightarrow \infty
\end{equation}
Here $\delta$ is the discrete delta-function (equal to one at the origin and zero elsewhere)
and $G(\mathbf{r}; k) = H_0^{(2)} (kr)$ is the continuous Green function, $H_0^{(2)}$ being the
Hankel function.

Without getting into the mathematical theory of lattice Green functions (see \cite{Martinsson02,Martinsson-PhD02}
and \sectref{sec:Lattice-Green}), let us note some features critical for our purposes:
\begin{itemize}
  \item
  In contrast with its continuous counterpart $G$, the discrete Green function
  \eqnref{eqn:Lg-eq-delta} is bounded everywhere, including the origin.
  \item The discrete Green function differs significantly from the continuous one
  only within a spatial window of several grid layers around the origin.
  Therefore only a relatively small amount of information
  needs to be stored -- namely, the values of the Green function within that window.
  This data can be precomputed for any given value of $kh$ and for each linear
  medium in a given problem.
\end{itemize}

\begin{figure}
\centering
  \includegraphics[width=\figurewidth]{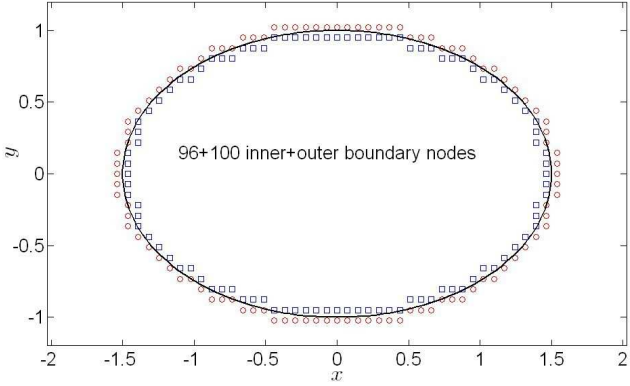}\\
  \caption{Discrete boundary $\gamma$ with 196 nodes.
  Squares: $\gamma_{\mathrm{in}}$; circles: $\gamma_{\mathrm{out}}$.}
  \label{fig:domain-bdry-gamma}
\end{figure}

The discrete boundary of the scatterer can be defined in a natural way follwoing
Ryaben'kii \cite{Ryabenkii02}.
Each grid node $m$ with discrete coordinates $(m_x, m_y)$ has four immediate neighbors from the respective five-point
stencil in the difference scheme \eqnref{eqn:5-pt-schem-E-mode-const-k}. If node $m$
lies \emph{inside} the scatterer but some of his neighbors are \emph{outside}, this grid node
will be said to belong to the discrete inner boundary $\gamma_{\mathrm{in}}$.
Likewise, if the central node $m$ of the stencil lies outside the scatterer but at least one of its neighbors
is inside, this node is said to belong to the discrete outer boundary $\gamma_{\mathrm{out}}$.
The complete discrete boundary consists of two layers:
$\gamma \equiv \gamma_{\mathrm{in}} \cup \gamma_{\mathrm{out}}$, \figref{fig:domain-bdry-gamma}.
(For larger multipoint stencils, the discrete boundary can be composed of several layers.)
The number of nodes on the boundary is $n_\gamma = n_{\gamma, \mathrm{in}} + n_{\gamma, \mathrm{out}}$.
These nodes can be referred to by pairs of indexes $(m_x, m_y)$ or, alternatively,
by some global numbers from 1 to $n_\gamma$. The order of this numbering makes no
principal difference but may slightly affect the practical implementation of the method.

%
\subsection{Boundary Sources}
%
The critical step is to express the lattice-based field in terms of fictitious discrete sources
$f$ that are nonzero only on the discrete boundary $\gamma$. For the inner boundary,
\begin{equation}\label{eqn:E-gammain-eq-f-star-g}
    E(\bfm) = [f * g(\cdot, \cdot; \kcyl)](\bfm),
    ~~ \bfm \equiv (m_x, m_y) \in \gamma_{\mathrm{in}}
\end{equation}
For the outer boundary,
%
\begin{equation}\label{eqn:E-gammaout-eq-f-star-g}
    E(\bfm) = E_{\mathrm{inc}}(\bfm) + [f * g(\cdot, \cdot; k_{\mathrm{out}})](\bfm),
    ~~ \bfm \in \gamma_{\mathrm{out}}
\end{equation}
The discrete convolution in the equations above is defined in the usual way as
\begin{equation}\label{eqn:discrete-convolution}
    [f * g](\bfn) ~\equiv~
    \sum_{\bfm \in \gamma} f(\bfm) g(\bfn - \bfm)
\end{equation}
Note that for the nodes on each side of the boundary the field is described
via the respective discrete Green function.

The auxiliary sources $f$ need not have
a direct physical interpretation, although ultimately they are indirectly related
to the equivalent electric and magnetic surface currents of traditional boundary integral methods
\cite{Peterson97,Gibson08}. However, the fields derived from these sources are physical.
The convolutions in equations \eqnref{eqn:E-gammain-eq-f-star-g}, \eqnref{eqn:E-gammaout-eq-f-star-g}
can be interpreted as (discrete) scattered fields.

It can be shown that any discrete field satisfying the FD wave equation on the lattice
can indeed be expressed via convolution with some fictitious boundary sources as stipulated above,
except possibly for special resonance cases (see Appendix).
%
\subsection{Boundary Difference Equations}
%
By construction, the electric fields defined by \eqnref{eqn:E-gammain-eq-f-star-g},
\eqnref{eqn:E-gammaout-eq-f-star-g} satisfy the respective wave equation on each side
of the boundary. What remains to be done then is to impose the boundary conditions; this will
lead to a system of equations from which the sources can be found.

To this end, one may use another difference scheme, $\mathcal{S}$, that approximates
the boundary conditions; we shall call it a ``boundary test scheme.'' The simplest example
is the five-point scheme
$$
    \mathcal{S}_5 E ~\equiv~
    E(m_x-1, m_y) + E(m_x+1, m_y) - 4 E(m_x, m_y)
$$
\begin{equation}\label{eqn:5-pt-schem-E-mode-var-k}
    + E(m_x, m_y-1) + E(m_x, m_y+1) + k^2(\bfm) h^2 E(\bfm) ~=~ 0
\end{equation}
In this second-order scheme, the value of $k$ is taken at the midpoint of the stencil.
A more accurate alternative is the nine-point FLAME (Flexible Local Approximation MEthod)
proposed in \cite{Tsukerman05,Tsukerman06,Tsukerman-book07}). Both types of schemes
are used in the numerical examples of \sectref{sec:Num-results}. The FLAME coefficients
$\mathcal{S}_9$ are computed as the nullspace of a matrix comprising the nodal
values of a set of basis functions on the stencil (\sectref{sec:FLAME} and
\cite{Tsukerman05,Tsukerman06,Tsukerman-book07}).

Applying a given boundary test scheme $\mathcal{S}$ on  $\gamma$ to fields
\eqnref{eqn:E-gammain-eq-f-star-g}, \eqnref{eqn:E-gammaout-eq-f-star-g},
one obtains a system of \emph{boundary-difference equations} of the form
\begin{equation}\label{eqn:F-f-star-g}
    \mathcal{S}^{(n)} [f * g(\cdot, \cdot; k(\bfn)](\bfn)
    ~=~ - \mathcal{S}^{(n)} E_{\mathrm{inc},h};
    ~~~ \bfn \equiv (n_x, n_y) \in \gamma
\end{equation}
where $E_{\mathrm{inc},h}$ is the discrete version of the incident field (i.e. its values on the lattice).
The superscript $(n)$ indicates that different schemes with different coefficients
can in principle be used over different stencils.

More explicitly, denoting the coefficients of the boundary test scheme $\mathcal{S}^{(n)}$ with $s_\alpha^{(n)}$
(where index $\alpha$ runs over nodes $\alpha$ over the grid stencil centered at node $n$),
one can write the boundary equation \eqnref{eqn:F-f-star-g} as
\begin{equation}\label{eqn:Af-eq-q}
    A f ~=~ q
\end{equation}
where $A$ is an $n_\gamma \times n_\gamma$ matrix with the entries
$$
   A_{nm} ~=~ \sum_\alpha s_\alpha^{(n)} g(\bfn - \bfm; k)
$$
and
$$
   q_n ~=~ - \sum_\alpha s_\alpha^{(n)} E_{\mathrm{inc},h}^{(n, \alpha)}
$$
The meaning of the terms above is as follows:
\begin{itemize}
  \item $n$, $m$ are the global numbers\footnote{Not to be confused with
  the Euclidean lengths of $\bfn = (n_x, n_y)$ and $\bfm = (m_x, m_y)$;
  these lengths are irrelevant and never appear in our analysis.}
  ($1 \leq n,m \leq n_\gamma$) of nodes
  $\bfn = (n_x, n_y)$ and $\bfm = (m_x, m_y)$ on the discrete boundary $\gamma$.
  \item $s_\alpha^{(n)}$ are the coefficients of the boundary test scheme corresponding to node $n$.
  (In principle, different schemes could be used at different nodes. One may even envision an adaptive
  procedure where the order of the scheme will vary in accordance with local accuracy estimates.)
  \item $k = \kcyl$ if node $n$ is on the inner boundary $\gamma_{\mathrm{in}}$ and
  $k = k_{\mathrm{out}}$ if it is on the outer boundary $\gamma_{\mathrm{out}}$.
  \item $E_{\mathrm{inc},h}^{(n, \alpha)}$ is the value of the incident field
  at node $\alpha$ of stencil n.
\end{itemize}

We shall call the numerical procedure leading to \eqnref{eqn:Af-eq-q}
the \emph{boundary difference method (BDM)}.
%
\section{The Lattice Green Function}
\label{sec:Lattice-Green}
%
As evident from the description of the BDM, lattice Green's functions
play a central role in it and must be computed accurately. There are at least two
general ways to do so: Fourier analysis and finite difference solutions.
A detailed exposition for the Laplace equation has been given by Martinsson \& Rodin \cite{Martinsson-PhD02,Martinsson02,Martinsson09}. Similar ideas can be immediately applied
to the wave equation as well, although a more elaborate analysis would be desirable
in the future.

Applying Fourier transform $\mathcal{F}$ (discrete physical space $\rightarrow$
continuous reciprocal space) to the difference equation \eqnref{eqn:Lg-eq-delta}
with the five-point operator $\mathcal{L}$ \eqnref{eqn:5-pt-schem-E-mode-const-k}, one obtains
$$
    \mathcal{F} \{ \mathcal{L} g \} \equiv
    (\exp(j \kappa_x) + \exp(-j \kappa_x) + \exp(j \kappa_y) + \exp(-j \kappa_y)
$$
$$
   - 4 + k^2 h^2) \mathcal{F} \{ g \} ~=~ 1
$$
where $ \kappa_x$, $ \kappa_y$ are the Fourier parameters in the square $[-\pi, \pi]^2$.


The inverse Fourier transform may then serve as a staring point for an asymptotic analysis
similar to Martinsson's \cite{Martinsson02,Martinsson-PhD02} and for practical computation
of Green's function $g$. However, this Fourier analysis is quite involved and must be
performed with great care, especially in 2D where Green's functions decay slowly
and regularization of Fourier integrals is necessary \cite{Martinsson02,Martinsson-PhD02}.
For the purposes of this paper, a more straightforward route
is sufficient. The finite difference problem \eqnref{eqn:Lg-eq-delta} for the Green function
can be solved directly, with the boundary condition \eqnref{eqn:g-eq-G} imposed
on the boundary of a large enough square $[-M, M]^2$. This can be done
efficiently with fast Fourier transforms over the square, but the computational
cost in 2D is so moderate that any other reasonable solver can be applied.
Obviously, one can also take advantage of the symmetries to reduce the size of the computational problem.

\begin{figure}
\centering
  \includegraphics[width=\figurewidth]{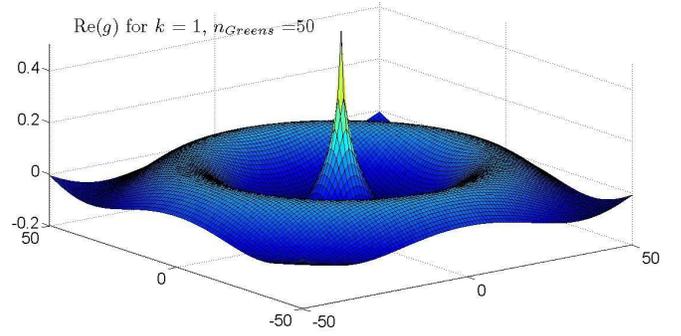}\\
  \caption{Re($g$) for $k = 1$, $h = 1/7$, $M = 50$.
  Note that the discrete green function is nonsingular everywhere; in fact,
  its magnitude in this example is quite moderate.}
  \label{fig:Re-g-k1-nGreens50}
\end{figure}

\begin{figure}
\centering
  \includegraphics[width=\figurewidth]{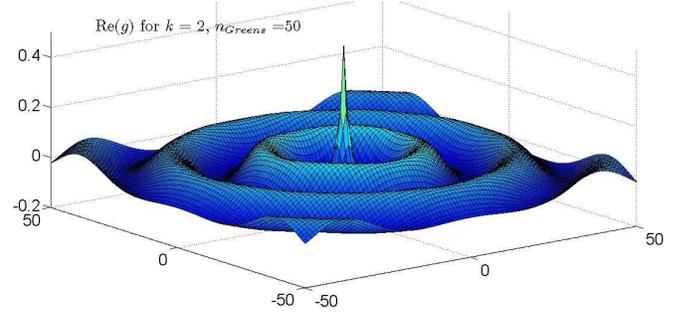}\\
  \caption{Re($g$) for $k = 2$, $h = 1/7$, $M = 50$.}
  \label{fig:Re-g-k2-nGreens50}
\end{figure}

The following plots illustrate the behavior of the lattice Green function and its computation.
All of the plots were generated for the grid size $h = 1/7$ as an example.
Surface plots of the real part of $g$ for wavenumbers $k = 1$ and $k = 2$
are shown in Figs. \ref{fig:Re-g-k1-nGreens50}
and \ref{fig:Re-g-k2-nGreens50}, respectively; Green's function was computed
in the spatial window $[-M, M]^2$ with $M = 50$.

\begin{figure}
\centering
  \includegraphics[width=\figurewidth]{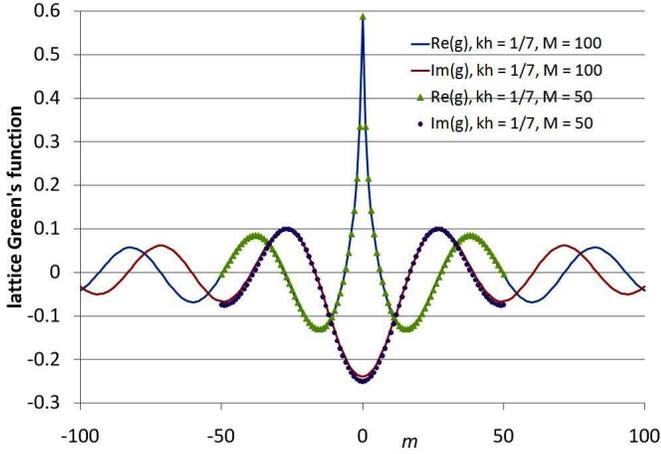}\\
  \caption{Lattice Green's function $g(m_x, m_y)$ vs $m_x$ for $kh = 1/7$,
  $m_y = 0$. The results for two different values of $M$,
  $M = 50$ and $M = 100$, are close.}
  \label{fig:g-vs-x-kh0142857-M50-100}
\end{figure}

\figref{fig:g-vs-x-kh0142857-M50-100} demonstrates that the size $M$ of the window
need not be too large. Indeed, lattice Green's functions for $M = 50$ and
$M = 100$ are quite close. The numerical experiments reported in the following
section were performed with $M = 50$. Even assuming an overkill value
$M = 100$ and 10 different materials in a given practical problem,
one ends up with less than 1 MB of data to be stored. In 3D, if one
takes advantage of the symmetries of $g$, the memory requirements are still reasonable,
even for vector fields and dyadic Green's functions, except for problems where
the number of different materials is unusually large.

%
%
\section{Numerical Simulation}
\label{sec:Num-results}
%
\subsection{FLAME}
\label{sec:FLAME}
The theory, implementation and various applications of FLAME have been discussed in a number
of previous publications \cite{Tsukerman05,Tsukerman06,Tsukerman-book07,Pinheiro07,Pinheiro09,
Tsukerman-JOPA09,Tsukerman-JCP10}, and therefore only a brief summary is given here.

FLAME replaces the usual Taylor expansions, the key tool of standard finite-difference
analysis, with much more accurate local (quasi-)analytical approximations of the solution. Such
approximations can be obtained, for example, via cylindrical or
spherical harmonics, plane waves, etc. Since the local behavior of
the field is ``built into'' the difference scheme, the accuracy
often improves dramatically. FLAME has already been applied to the simulation of
colloidal and plasmonic particles
\cite{Tsukerman05,Tsukerman06,Tsukerman-book07}, negative-index
materials \cite{Cajko08}, the computation of
Bloch bands of photonic crystals \cite{Tsukerman-PBG08}, including complex
bands for plasmonic systems and other dispersive media \cite{Tsukerman-PBG08,Tsukerman-JOPA09}.

For the model problems in this paper, local analytical bases for FLAME are
available via Bessel / Hankel functions. More specifically, in the vicinity of a
dielectric cylinder \emph{with a circular cross-section} centered for convenience at the origin of a polar
coordinate system $(r, \phi)$, these approximating functions -- the
FLAME basis $\psi_\alpha^{(i)}$ -- are
\cite{Tsukerman05,Tsukerman06,Tsukerman-book07}
$$
    \psi_\alpha^{(i)} ~=~  a_l^{(i)} J_l (k_\mathrm{cyl} r) \exp(\iu l \phi) ,
    ~~ r \le r_{\mathrm{cyl}}
$$
$$
    \psi_\alpha^{(i)} ~=~ [c_l^{(i)} J_l(k_{\rm air} r) + H_l^{(2)}
    (k_\mathrm{out} r)]  \exp(\iu l \phi), ~~ r > r_{\mathrm{cyl}}
$$
where $J_l$ is the Bessel function of order $l$, $H_l^{(2)}$ is the Hankel function of
the second kind, and $a_l^{(i)}$, $c_l^{(i)}$ are coefficients to
be determined. These coefficients are found via the standard
conditions on the boundary of the cylinder
\cite{Tsukerman05,Tsukerman-book07,Tsukerman-PBG08}. Index $i$ runs
over all grid stencils where the FLAME scheme is generated, while
index $\alpha$ runs over all basis functions in a given stencil $i$.

In this paper, the 9-point $(3 \times 3)$ stencil with a grid size
$h$ is used and $1 \leq \alpha \leq 8$. The eight basis functions $\psi$ are
obtained by retaining the monopole harmonic ($l$ = 0), two harmonics
of orders $|l| = 1, 2, 3$ (i.e. dipole, quadrupole and octupole), and
one of the harmonics of order $|l| = 4$. This set of basis functions
produces a nine-point scheme as the null vector of the respective
matrix of nodal values \cite{Tsukerman05,Tsukerman06}.

For the test problem with {an \emph{elliptical} cylinder (\sectref{sec:Num-results-ellipse}),
it is still possible to use the same Bessel-Hankel basis functions in FLAME. Toward this end, a piece of the ellipse
straddled by a given grid stencil is approximated by its osculating circle (with the radius
equal to the radius of the curvature of the ellipse at a given point on its boundary).
While this approach for constructing FLAME bases is fairly straightforward,
it has never been used previously. (In the past, the primary motivation
was to apply FLAME on very coarse grids that carry almost no information
about the shape of the boundary \cite{Tsukerman05,Tsukerman06,Tsukerman-book07}.)

For the ellipse, the osculating circle can easily be found analytically;
for more complicated boundaries, the curvature could approximately be evaluated
numerically -- for example, as the best local fit to a piece of the discrete boundary $\gamma$.
In yet more complex cases -- especially in 3D where there are two radii of curvature --
one could use piecewise-planar approximations and Fresnel-formula FLAME bases
\cite{Cajko08}.

%
\subsection{Circular cylinder}\label{sec:Num-results-circular}
%
For verification, let us first consider a circular scattering cylinder,
as in this case a well-known analytical solution via cylindrical harmonics
exists. In all numerical experiments below,
the $E$ mode was considered. The wavenumber for the incident wave
was normalized at $k_{\mathrm{out}} = 1$; the wavenumber for the scatterer
was taken as $k_{\mathrm{cyl}} = 2$ (i.e. $\epsilon_{\mathrm{cyl}} = 4$). The incident plane
wave propagates in the positive $x$ direction. The color plot of the electric field
in BDM with $n_\gamma = 460$ is shown in \figref{fig:ReE-kcyl2-n40-ngamma460-nGreens50}.

\begin{figure}
\centering
  \includegraphics[width=\figurewidth]{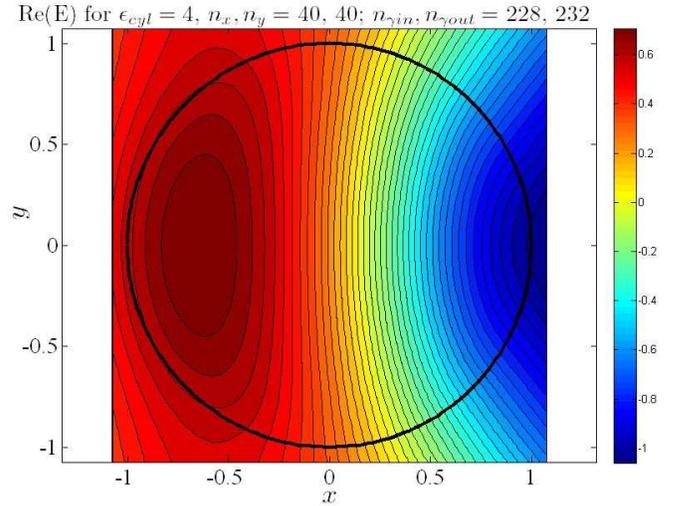}\\
  \caption{Color plot of the $E$ field for a circular cylinder with $\epsilon_{\mathrm{cyl}} = 4$.
  BDM with $n_\gamma = 460$.}
  \label{fig:ReE-kcyl2-n40-ngamma460-nGreens50}
\end{figure}

The numerical error as a function of the BDM grid size $h$ is plotted in \figref{fig:error-vs-h-scattering-cylinder}
for two boundary test schemes $\mathcal{S}$: the standard five-point scheme and the nine-point FLAME
(these schemes were briefly described in the previous sections). The dashed line in the figure serves only
as a visual aid indicating the second order convergence of the method for both schemes.
Not surprisingly, the numerical error for FLAME is about an order of magnitude lower than
that of the five-point scheme. However, the order of convergence is still limited
by the second-order five-point difference scheme used to compute the discrete Green function
(\sectref{sec:Lattice-Green}).
The relative error was calculated as $\|E_{\mathrm{BDM}} - E_{\mathrm{exact}}\| / \| E_{\mathrm{exact}}\|$,
where $E_{\mathrm{BDM}}$ and $E_{\mathrm{exact}}$ are the numerical and the quasi-exact
solutions on the grid, respectively; the norms are Euclidean. The quasi-exact solution
was computed via the standard expansion into cylindrical harmonics up to order 50.

\begin{figure}
\centering
  \includegraphics[width=\figurewidth]{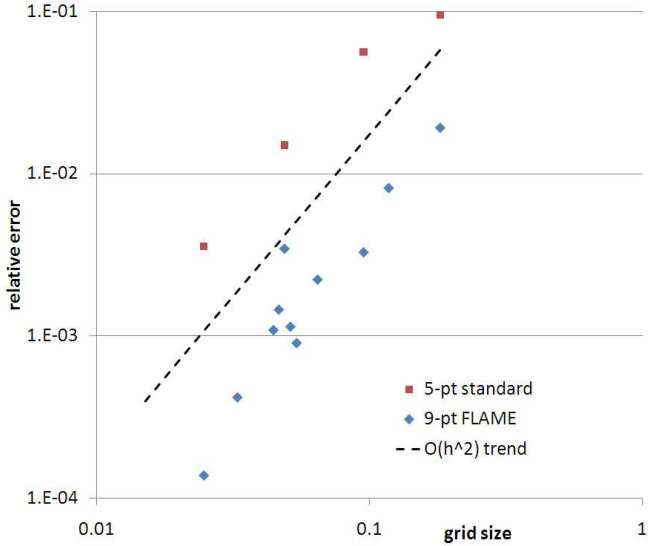}\\
  \caption{The relative error in BDM as a function of grid size. Discrete boundary $\gamma$ with 196 nodes.
  Quadratic convergence, commensurate with the order of the scheme for the lattice Green function, is observed.
  The FLAME results are about an order of magnitude more accurate than for the standard five-point scheme,
  even around $h \sim 0.05$ where the FLAME data points exhibit some scatter.}
  \label{fig:error-vs-h-scattering-cylinder}
\end{figure}
%
%
\subsection{Elliptical Cylinder}
\label{sec:Num-results-ellipse}
%
The simulations have been repeated for an elliptical cylinder,
with the same physical parameters as above, and with the ratio of the axes 1.5 : 1.
\figref{fig:Re-E-kcyl2-n40-ngamma196-nGreens40-ellipse}
is a color plot of the real part of the electric field obtained with BDM,
9-point FLAME scheme, discrete boundary $\gamma$ with 196 nodes.
FLAME was generated as described at the end of
\sectref{sec:FLAME}: by locally approximating a piece of the elliptic boundary
with a circle and using the respective Bessel / Hankel bases.

\begin{figure}
\centering
  \includegraphics[width=\figurewidth]{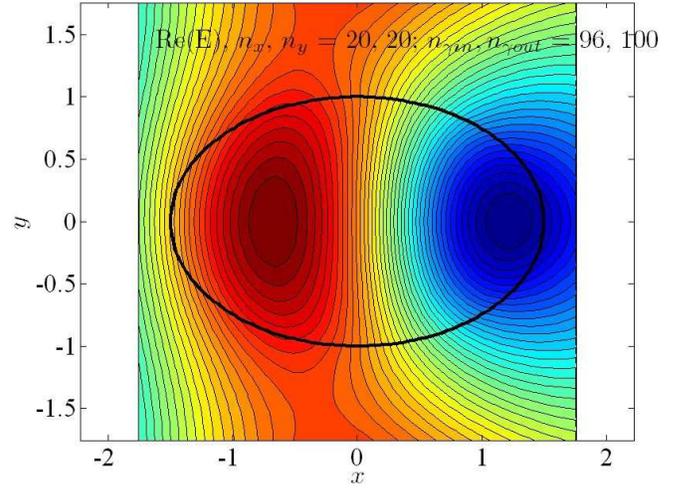}\\
  \caption{Re($E$) obtained with BDM, 9-point FLAME scheme.
  Discrete boundary $\gamma$ with 196 nodes.}
  \label{fig:Re-E-kcyl2-n40-ngamma196-nGreens40-ellipse}
\end{figure}

\figref{fig:Re-E-vs-coordinate-ellipse-nGreens40} demonstrates that the field distributions
obtained with different methods are in a very good agreement. Plotted in the figure is
the real part of the electric field vs. $x$ (at $y = 0$) and vs $y$ (at $x = 0$). The imaginary parts
are not plotted, but agree with the theory equally well. Nine-point FLAME and standard five-point
schemes were applied on several grids with sizes
$h = r_{cyl} / (n + \frac12)$; results for $n = 10$ and $ n = 20$ are shown.

\begin{figure}
\centering
  \includegraphics[width=\figurewidth]{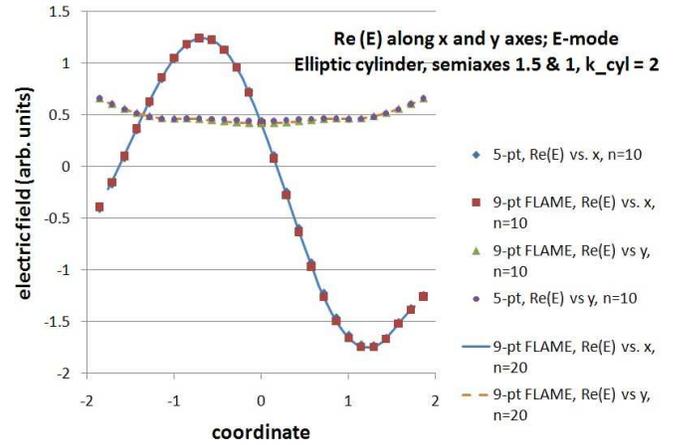}\\
  \caption{The numerical results for different cases
  are seen to be in a very good agreement. The real part of the electric field is plotted;
  the agreement for the imaginary part is similar.
  Discrete boundary $\gamma$ with 196 nodes.}
  \label{fig:Re-E-vs-coordinate-ellipse-nGreens40}
\end{figure}
%
\section{Discussion and Conclusion}
\label{sec:Discussion}
%
\subsection{Summary}
The boundary difference method described and implemented in this paper for
wave scattering avoids the singularities inherent in traditional boundary integral methods.
This is accomplished by reversing the sequence of stages in the procedure.
Traditionally, the differential equations are first reduced to
boundary integrals with respect to equivalent sources on the boundary and then discretized;
the kernels of the underlying integral equations are singular due to the infinite
self-fields of concentrated sources.

In BDM, the differential problem is first discretized on a regular grid to obtain a finite-difference
approximation that is then reduced to a boundary difference equation with respect to auxiliary
sources on the discrete boundary. The field of these sources can be expressed by convolution
with the discrete Green function that, unlike its continuous counterpart, is finite at all points.
Thus no singularities ever arise.

Technically, the underlying grid is infinite. The computational procedure, however,
involves only the boundary nodes of the grid and a finite spatial window where the discrete Green function
is precomputed, which can be done once and for all for a given set of parameters.

The validity of BDM has been demonstrated using 2D scattering from dielectric cylinders
as a model problem. Convergence of the method as a function of the grid size has been established and is
commensurate with the order of finite-difference schemes used.
%
\subsection{Trade-offs}\label{sec:Trade-offs}
%
Since the proposed approach has common features with the traditional integral equation methods,
some of the usual trade-offs between differential and integral techniques \cite{Konrad93}
apply. The differential methods lead to sparse matrices, whereas the boundary
methods produce dense ones. This drawback can be partly alleviated via fast multipole
acceleration \cite{Greengard87,Martinsson-PhD02,Ying04,Martinsson09,Fong09}. Its use in conjunction
with BDM is relatively straightforward. Indeed, FMM relies on a recursive splitting of the solution
into near- and far-field components. The far field in BDM is essentially the same
as in the continuous problem, by construction of the discrete Green function; see
\eqnref{eqn:g-eq-G}. It is only in the near field that discrete and continuous Green functions
may differ significantly, but this makes little difference in FMM algorithms
because the near-field contribution is computed directly.

As already emphasized, BDM completely dispenses with singular integral kernels,
an inherent drawback of integral methods. The price to pay for that is the need
to precompute discrete Green functions. In practice, this price can be
expected to be modest, because the number of different materials in any given problem
is limited and the computation involves a relatively small number of grid layers
around Green's point source. In any event, this computational overhead is independent of the size of the
problem being solved.

For unbounded problems, differential methods such as FEM and FD require artificial domain truncation
with absorbing boundary conditions or matched layers. No such truncation is needed in boundary methods.
At the same time, differential methods are generally better suited for nonlinear problems
that call for volume discretization, in which case the boundary methods
usually lose their effectiveness.

The key source of numerical errors in traditional BEM is approximation of singular integrals
(typically, by piecewise-polynomial functions of low order, including piecewise-constant
approximations in the simplest case). In BDM, the error is due to the finite-difference approximation
of the boundary conditions and of the lattice Green function. If the order of these approximations
is increased, the overall numerical error of the method can be reduced accordingly.
%
\subsection{Generalizations and Future Directions}\label{sec:Future-directions}
%
Boundary difference schemes developed in this paper lend themselves to generalization
in quite a natural way.
Unlike traditional boundary integral methods, BDM is automatic, in the sense
that it does not require the suitable sets of equivalent boundary sources
(electric or magnetic surface currents, surface charges, etc.) and the respective
equations to be worked out in advance. Instead, one introduces discrete boundary
sources that need not even have a specific physical meaning; but once computed, they can be used to find physical fields
by convolution with Green's functions on the lattice.
In particular, the $H$-mode (TE- or $p$-mode) of electromagnetic wave
scattering is treated in BDM exactly the same way as the $E$-mode.
(As a side note, for the $H$-mode the classical five-point control volume scheme would only be
of order one at the boundary, but that is a feature of that scheme, not of BDM as a whole.)

Further, extension to 3D vector problems is also conceptually straightforward,
although clearly algorithmic challenges do arise. This line of research
is currently being pursued.

The boundary difference method does in general require a spatially uniform grid.
Although this grid is ``virtual,'' in the sense that the actual computation
involves only the nodes on the discrete boundary and not the volume nodes,
the uniformity of the grid may still be a limiting factor in some problems.
However, if several scatterers are present and well-separated (in practice,
by at least a few grid layers), then each of them may be meshed separately.
Indeed, in that case the interactions between different scatterers are numerically
in the far field, where the continuous Green function can be used as a good proxy
for the discrete one.

Finally, the method is not limited to electromagnetics and can be extended to
other classes of linear problems, including acoustics and elasticity.
It may even be applied to micro-, nano- and molecular-scale models
on a discrete lattice (e.g. Haq \etal \cite{Haq06}), when continuous equations may not even be available.
%
\section*{Appendix: Representation of the Field via Discrete Sources}
%
Let us show that any discrete field on the boundary $\gamma$ can be represented
via convolution of the Green functions with some auxiliary sources on the same boundary,
except possibly for some special cases of interior resonance.
More precisely, let $E_{sh} = E_h - E_{\mathrm{inc,h}}$ be the scattered component of a lattice field
$E_h$ that satisfies the discretized wave equation both inside and outside the scatterer.
Further, let $E_{sh, \gamma}$ represent the values of $E_{sh}$
on the discrete boundary $\gamma$. We intend to show that
\begin{equation}\label{eqn:Es-eq-f-star-g}
   E_{sh, \gamma} (\bfm) ~=~ [f * g(\cdot, \cdot; k(\bfm))](\bfm)
\end{equation}
for some source $f$ on $\gamma$.

The discrete convolution in \eqnref{eqn:Es-eq-f-star-g} can be viewed as a linear operator that maps
functions $f$ in $R^{n_\gamma}$ to fields $E_{sh, \gamma}$, also in $R^{n_\gamma}$. It is then
sufficient to demonstrate that this operator is nonsingular or, equivalently, that an identically zero field
on the discrete boundary can be produced only by zero sources on that boundary.

Let us thus assume that $E_{sh, \gamma}$ is identically zero. Then $E_{sh}$ must be zero, too,
\emph{everywhere in the outside region}. This is true because, by its construction as convolution
with sources only on $\gamma$, this field satisfies the homogeneous difference equation in the
outside region and also, by assumption, the Dirichlet conditions for it on $\gamma_{\mathrm{out}}$ are zero.
Similar considerations hold for $E_{sh}$ in the inside region away from the interior resonance,
as long as $\kcyl$ is not an eigenvalue of the wave problem inside the scatterer,
with zero Dirichlet conditions. Thus the convolution in \eqnref{eqn:Es-eq-f-star-g}
must be identically zero \emph{on the whole lattice}, from which it immediately follows
(e.g. via Fourier transforms) that $f = 0$. \hskip 0.2in $\therefore$
%
\section*{Acknowledgment}
%
I am very grateful to S. V. Tsynkov for informative and illuminating discussions
that helped to catalyze the research reported in the paper. I also thank
P.-G. Martinsson and G.~Rodin for helpful comments.

\end{document}